\documentclass[useAMS,usenatbib]{pasa}

\usepackage{graphicx}

\newcommand{\HI}{{\sc{H}{I}}}
\newcommand{\msun}{$M_\odot$}
\newcommand{\etal}{{et~al.}}
\newcommand{\mhi}{$M_{HI}$}
\newcommand{\kms}{~km\,s$^{-1}$}

\title[\HI\ and Galaxy Formation in Loose Groups]{\HI\ and Galaxy Formation 
in Loose Groups}
\author[D.J. Pisano]{D.J. Pisano,$^{1,2,3}$}
\affil{
$^1$CSIRO ATNF, P.O. Box 76, Epping, NSW 1710, Australia\\
$^2$Bolton Fellow, NSF MPS Distinguished International Postdoctoral Fellow\\
$^3$dj.pisano@csiro.au}

\begin{document}

\maketitle

\label{firstpage}

\begin{abstract}
Models of hierarchical galaxy formation predict that large numbers of
low-mass, dark matter halos remain around galaxies today.   These models 
predict an order of magnitude more halos than
observed stellar satellites in the Local Group.  One possible solution to
this discrepancy is that the high-velocity clouds (HVCs) around the
Milky Way may be  associated with the excess dark matter halos and be
the gaseous remnants of the galaxy formation process.  If this is the
case, then analogs to the HVCs should be visible in other groups.  In
this paper, we review the observations of \HI\ clouds lacking stars
around other galaxies and in groups and present early results from
our \HI\ survey of loose groups analogous to the Local Group and its
implications for the nature of HVCs and galaxy formation.
\end{abstract}

\begin{keywords}
Local Group -- intergalactic medium -- galaxies: dwarf -- galaxies: formation
\end{keywords}

\section{Introduction}
Current models of hierarchical galaxy formation predict that galaxies
form via the accretion of smaller lumps of gas, stars, and dark matter
(e.g. Silk \& Norman 1981; Kauffman, White, \& Guiderdoni 1993; Cole
\etal\ 1994).  Recent simulations of this process assuming a
lambda-dominated cold dark matter (CDM) universe uniformly reveal the
presence of large numbers of low-mass dark matter halos persisting
around larger galaxies into the present day (e.g. Klypin \etal\ 1999,
Moore et al. 1999).  It remains unclear if these dark matter halos are
filled with gas and/or stars and can be associated with dwarf galaxies
and/or \HI\ clouds or if they lack the mass to retain any baryons.
This raises the  question:  do we see such  gaseous remnants of galaxy
formation in \HI\ emission around galaxies today?

There have been many detections of \HI\ clouds around nearby galaxies,
many of which have been hypothesized to be primordial gas associated
with galaxy formation.  No \HI, however, has been unambiguously
associated with galaxy formation.  Its origin can be more readily
attributed to tidal interactions, galactic fountains, or galaxy
accretion.

NGC 4449 is an irregular galaxy with two counter-rotating gas
complexes and extended \HI\ distributed in large clouds, arms, and
streamers  (Hunter \etal\ 1998).  While this \HI\ could be explained
as infalling primordial gas, the distribution and kinematics of the
gas are perhaps more  simply explained as resulting from a tidal
interaction with a nearby dwarf galaxy (Theis \& Kohle 2001).

In NGC 6946, many clouds of \HI\ are seen moving at velocities
inconsistent with the rotation of the galaxy (Kamphuis \& Sancisi
1993).   This high-velocity gas is widely associated with holes in the
\HI\ disk,  and may be explained as gas that was ejected via a
galactic fountain  powered by supernovae in the galaxy.  An
alternative explanation is that these clouds are infalling primordial
material which has punched holes in the galaxy (Kamphuis \& Sancisi
1993).

There are many other cases of \HI\ seen outside of the main body of
galaxies which have less certain origins.   IC 10 (Wilcots \& Miller
1998) and NGC 925 (Pisano, Wilcots, \& Elmegreen 1998) are two
examples of galaxies in groups of galaxies which have no stellar
companions within $\sim$100 kpc, but both have  \HI\ clouds of
$\sim$10$^7$\msun\ within a few tens of kiloparsecs.  These  clouds
could be the remains of dwarf galaxies which have been torn apart by
the larger galaxy, or they could be tidal debris from an ancient
interaction, or these clouds could be primordial \HI\ gas falling into
these galaxies for the  first time contributing to the ongoing
assembly of these galaxies.

Finally, surrounding our own galaxy are the high-velocity clouds
(HVCs):  \HI\ clouds which lack stars and are moving at velocities
inconsistent with  Galactic rotation.   Because of this, we can not
infer their distances or their masses  (see Wakker \& van Woerden 1997
for a review).   HVCs most likely represent a variety of phenomena.
Some HVCs are probably related to a galactic fountain (Shapiro \&
Field 1976; Bregman 1980) and are located in the lower Galactic halo.
Other HVCs are certainly tidal in origin: the Magellanic Stream is the
most obvious of these features, formed via the  tidal interactions
between the Milky Way, Large Magellanic Cloud, and Small Magellanic
Cloud (e.g. Putman \etal\ 1998), with other HVCs potentially related
to other satellites, such as the Sagittarius dwarf (Putman \etal\
2004).  And some HVCs, such as Complex C, may be infalling primordial
gas (Wakker \etal\ 1999;  Tripp \etal\ 2003; cf. Gibson et al. 2001).
Finally, Blitz \etal\ (1999) and Braun \& Burton (1999) suggested that
HVCs are the debris from the formation of  the Local Group, and not
just the Milky Way, and are associated with dark  matter halos.  In
this scenarios HVCs are distributed throughout the Local  Group with
D$\sim$100 kpc -- 1 Mpc and \mhi$\sim$10$^{5-7}$\msun.

Klypin \etal\ (1999) and Moore \etal\ (1999) raised the issue that CDM
models  of galaxy formation predict that the Local Group should have
hundreds of small  dark matter halos, while the number of known, luminous
satellite galaxies is only  $\sim$20.  If HVCs are associated with
dark matter halos then this discrepancy,  the ``missing satellite''
problem, would be resolved.  Also at issue is if the Local Group is
somehow unique in this regard and other galaxies have a sufficient
number of satellites to match predictions.

Unfortunately, none of the \HI\ detections discussed above are clearly
associated with galaxy formation.  All of the detections may have
alternative  origins.  To better constrain the origin of such \HI\
clouds, a more systematic  search for them is required.  I discuss
here the early results of such an \HI\  survey of loose groups
analogous to the Local Group using the Parkes Multibeam  instrument
and done in collaboration with David Barnes (Melbourne), Brad Gibson
(Swinburne), Lister Staveley-Smith (ATNF), and Ken Freeman (ANU).
This survey will determine if there are massive analogs to HVCs in
other groups of galaxies associated with dark matter halos as
described above.   Furthermore, this survey will find gas-rich dwarf
galaxies and determine if the ``missing satellite'' problem is unique
to the Local Group or a ubiquitous problem in all groups.  Finally,
these observations will serve as a benchmark for the \HI\ properties
of galaxies in spiral-rich loose groups and how they compare to
galaxies in other environments.

\section{Survey Parameters \& Results}

To search for HVC analogs, to test models of galaxy formation, and to
better understand the \HI\ properties of groups like the Local Group,
we have surveyed six spiral-rich, loose groups in \HI\ 21 cm emission
using the Parkes Multibeam instrument (Pisano \etal\ 2004a,b).   A
loose group is a collection a few large galaxies and tens of smaller
ones, where the large galaxies are well-separated, of order a few
hundred kpc, over an area of $\sim$1 Mpc$^2$.  In contrast to compact
groups, such as Stephan's Quintet and those cataloged by Hickson
(1982) where interactions are a driving force, loose groups generally
have few interactions occurring.  We were particularly interested in
loose groups containing only large spiral galaxies; groups analogous
to the Local Group.  If HVCs are associated with the formation of the
Local Group, then they should be present in these groups as well.

The Parkes Multibeam observations were conducted in six separate
observing runs between October 2001 and June 2003.  The Multibeam
instrument was repeatedly scanned in right ascension and declination
over an area of $\sim$1 Mpc$^2 \equiv$25 square degrees with a velocity
coverage of $>$1500 \kms\ until an rms sensitivity of 5-8 mJy
beam$^{-1}$  per 3.3 \kms\ channel was reached.  This translates to a
\mhi\ sensitivity at the distances of these groups, 10.6$\,-\,$13.4 Mpc,
of 5.3 - 8.1$\times$10$^5$\msun\ per channel.  The Parkes data had
fake sources inserted before multiple double-blind searches by eye for
all sources were made.  All sources, not just new ones, were confirmed
with follow-up Australia Telescope Compact Array (ATCA) observations.
Based on our identification of fake sources, we determined that our
searches were nearly 100\% complete down to an integrated flux limit of 
ten times the rms noise times the square root of the number of channels (Pisano
\etal\ 2004a,b).  Only three groups have had their Parkes detections
confirmed with ATCA follow-up observations:   LGG 93, 180, \& 478.
In these three groups, all 20 previously optically-identified group
members were detected as well as seven new \HI-rich dwarf galaxies.
No \HI\ clouds without stars identified on the Digital Sky Survey were
found.

\section{Implications for High Velocity Clouds}

Because Pisano \etal\ (2004a) did not detect any HVC analogs in the
three  groups surveyed, we are unable to confirm the existence of
analogs of the type proposed by Blitz \etal\ (1999) and Braun \&
Burton (1999).  We can, however, place limits on the masses and,
hence, distances of the HVCs around the Milky Way if such objects are
ubiquitous in the group environment.  This has been done by Pisano
\etal\ (2004b).  Pisano \etal\ examine only the compact HVCs (CHVCs)
identified by Braun \& Burton (1999) as likely being associated with
dark matter halos.  This is because many of the other classes of
HVCs, such as the Magellanic Stream and large complexes, discussed by
Blitz \etal\ (1999) as being associated with dark matter have other,
more likely, origins as discussed above.  Pisano \etal\ assume that
CHVCs are distributed in a Gaussian manner about the Milky Way, and
ask for what parent population of CHVCs and what D$_{HWHM}$ for their
distribution would we expect to see {\it zero} analogs around
galaxies in the loose groups surveyed given the detection limits of
the observations.

\begin{figure}
\includegraphics[scale=0.4, angle=-90]{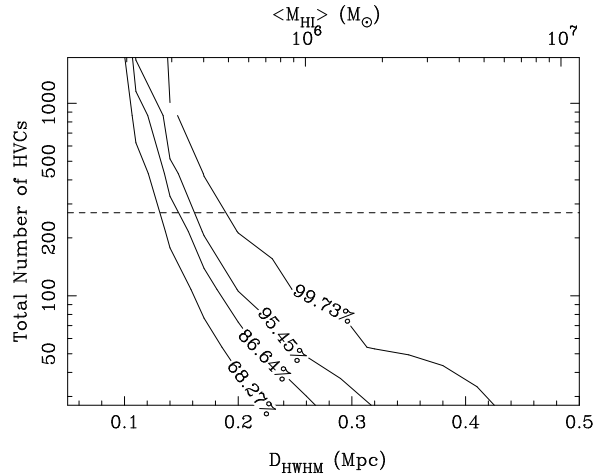}
\caption{A plot of the combined probability of zero detections in LGG
93, 180, \& 478  as a function of the number of CHVCs per group and
D$_{HWHM}$ (equivalent to the  average \mhi\ of a CHVC) for the
distribution of Milky Way CHVCs.  The dashed line  marks the number of
CHVCs identified around the Milky Way.  \label{hvc}}
\end{figure}

Figure~\ref{hvc} illustrates the combined constraints from the
observations of LGG 93, 180, \& 478 on the distances and population of
CHVCs around the Milky Way.  If the CHVC population in other groups
has the same properties and as those around the Milky Way, then at the
95\% confidence level, for 270 clouds, we see that CHVCs must be
clustered within 160 kpc of the Milky Way with an average \mhi\ of
$\le$4$\times$10$^5$\msun.  This is in good agreement with recent
limits derived by other authors using a variety of other methods
examining both Milky Way HVCs (e.g. de Heij \etal\ 2002) and
extragalactic analogs (e.g. Zwaan 2001) and makes the original Blitz
\etal\ (1999) and  Braun \& Burton (1999) models which place HVCs at
$\sim$1 Mpc extremely unlikely.  These limits imply that CHVCs are
more closely associated with individual galaxies, rather than groups
of galaxies, and that there is not a large reservoir of neutral
hydrogen, $\le$1$\times$10$^8$\msun, waiting to be accreted onto
galaxies like the Milky Way.  These observations do not rule out the
presence of a large reservoir of ionized gas, however.

\section{Comparison with models of Galaxy Formation}

\begin{figure}
\includegraphics[scale=0.4, angle=-90]{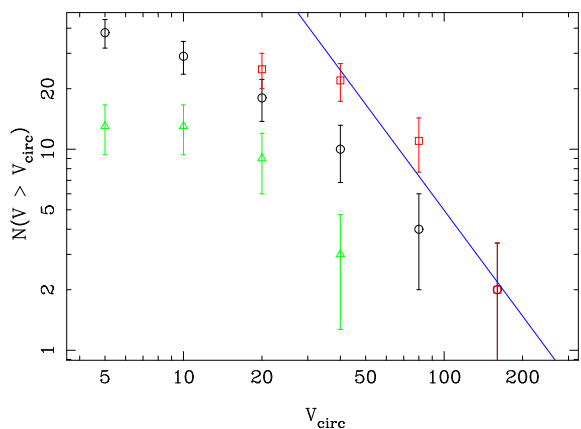}
\caption{A plot of the cumulative velocity distribution function for
the Local Group (black circles), our three loose groups (red squares),
and companions to isolated galaxies (green triangles).  Error
bars indicate 1$\sigma$ Gaussian errors.   The solid blue line is the 
prediction for CDM models of galaxy formation from  Klypin \etal\ (1999) with
arbitrary normalization. \label{cvdf}}
\end{figure}

The observations of loose groups by Pisano \etal\ (2004a) do not find
large numbers of \HI-rich galaxies that would correspond to the low mass
dark matter halos seen in the simulations by Klypin \etal\ (1999) or Moore
\etal\ (1999).  As such we can infer that the Local Group is not unique in 
its lack of low mass, luminous satellites as compared to the predictions of
CDM simulations.  Pisano \& Wilcots (2003) previously found this to be
the case for gas-rich companions to isolated galaxies.  These results are
illustrated in Figure~\ref{cvdf}. 

The cumulative velocity distribution functions for the Local Group,
loose groups, and isolated galaxies have roughly consistent slopes which are
inconsistent with the CDM models.  From this it is clear that our measurements
of luminous halos do not match CDM predictions and either an alternative
form of dark matter (such as warm dark matter, Col\'{\i}n \etal\ 1999)
or a mechanism suppressing the collapse of baryons into dark halos
(e.g. Tully \etal\ 2002) is needed to reconcile the observations with
models.  These results will be discussed in more detail in future
papers (Pisano \etal, in preparation).

\section*{Acknowledgements}

I would like to thank the organizers of the Tullyfest for inviting me to 
present this work at such an excellent meeting.  I would also like to thank 
the staff at Parkes and the ATCA for their wonderful support of these 
observations.  I would like to acknowledge generous support from both an ATNF 
Bolton Fellowship and NSF MPS Distinguished International Research Fellowship 
grant AST0104439.

\label{lastpage}

\end{document}